# A-15 type superconducting hydride La$_4$H$_{23}$: Nanograined structure with low strain, strong electron-phonon interaction, and moderate level of nonadiabaticity

Evgeny F. Talantsev[1,2] and Vasiliy V. Chistyakov[1,2]

[1]M.N. Miheev Institute of Metal Physics, Ural Branch, Russian Academy of Sciences, 18, S. Kovalevskoy St., Ekaterinburg, 620108, Russia
[2]NANOTECH Centre, Ural Federal University, 19 Mira St., Ekaterinburg, 620002, Russia

**Abstract**

For seven decades by A-15 superconductors we meant metallic A$_3$B alloys (where A is a transition metal, and B is groups IIIB and IVB element) discovered by Hardy and Hulm (*Phys. Rev.* **89**, 884 (1953)). Nb$_3$Ge exhibited the highest superconducting transition temperature, $T_c$ = 23 K, among these alloys. One of these alloys, Nb$_3$Sn, is primary material in modern applied superconductivity. Recently Guo *et al* (*arXiv*:2307.13067) extended the family of superconductors where the metallic ions arranged in the beta tungsten (A-15) sublattice by observation of $T_{c,zero}$ = 81 K in La$_4$H$_{23}$ phase compressed at $P$ = 118 GPa. Despite the La$_4$H$_{23}$ has much lower $T_c$ in comparison with near-room-temperature superconducting LaH$_{10}$ phase ($T_{c,zero}$ = 250 K at $P \sim$ 200 GPa) discovered by Drozdov *et al* (*Nature* **569**, 531 (2019)), the La$_4$H$_{23}$ holds the record high $T_c$ within A-15 family. Cross *et al* (*Phys. Rev. B* **109**, L020503 (2024)) confirmed the high-temperature superconductivity in the compressed La$_4$H$_{23}$. In this paper, we analyzed available experimental data measured in La$_4$H$_{23}$ and found that this superconductor exhibits nanograined structure, 5.5 nm $\leq D \leq$ 35 nm, low crystalline strain, $|\varepsilon| \leq$ 0.003, strong electron-phonon coupling interaction, 1.5 $\leq \lambda_{e-ph} \leq$ 2.55, and moderate level of the nonadiabaticity, 0.18 $\leq \Theta_D/T_F \leq$ 0.22 (where $\Theta_D$ is the Debye temperature, and $T_F$ is the Fermi temperature). We found that derived $\Theta_D/T_F$ and $T_c/T_F$ values for the La$_4$H$_{23}$ phase are similar to the ones in MgB$_2$, cuprates, pnictides, and near-room-temperature superconductors H$_3$S and LaH$_{10}$.

# A-15 type superconducting hydride La$_4$H$_{23}$: Nanograined structure with low strain, strong electron-phonon interaction, and moderate level of nonadiabaticity

## I. Introduction

In 1953 Hardy and Hulm[1] discovered that A$_3$B alloys (where A is one of the transition metals Ti, Zr, V, Nb, Ta, Cr, and Mo, and B is one element of groups IIIB and IVB, or the precious metals Os, Ir, Pt, and Au[2]) with A-15 lattice exhibit the superconducting transition temperature up to $T_c$ = 23 K (for Nb$_3$Ge[3]) and high values for low-temperature upper critical field, $B_{c2}$(4.2 K) ~ 37 T (for Nb$_3$Ge[4]). One of these alloys, Nb$_3$Sn, is primary material for superconducting wires in nearly all modern commercial magnetic systems, including magnetic systems for mega-science projects[5–37].

Primary physical reason why metallic hydrogen and hydrides are key materials in the quest of room temperature superconductivity can be understood based on two conclusions of the Bardeen-Cooper-Schrieffer theory[38] of the electron-phonon mediated superconductivity:

1. $T_c \cong 1.17 \times \Theta_D \times e^{-\frac{1}{\lambda_{e-ph}-\mu^*}}$ (1)

2. $T_c \propto \frac{1}{\sqrt{M}}$ (2)

where $\Theta_D$ is the Debye temperature, $\lambda_{e-ph}$ is the dimensionless electron–phonon interaction constant, $\mu^*$ is reduced electron–electron interaction constant (the Coulomb pseudopotential), $M$ is the mass of the metallic ion. Despite strict theory of the electron-phonon mediated superconductivity[39,40] and its later development[41–43] are complicated, simplified Eqs. 1,2 are very useful tool to understand primary physical idea for the quest[44–46] of room-temperature superconductivity, where the desirable parameters of the superconductor are:

1. high Debye temperature, $\Theta_D$,
2. strong electron-phonon interaction, $\lambda_{e-ph} \geq 1.5$, and $\lambda_{e-ph} \gg \mu^*$;
3. as low as possible, the metal ion mass, $M$.

Hypothetical metallic state of hydrogen[47] would satisfy these conditions. However, it is extremely difficult to create a metallic state in hydrogen in experiment[48–51]. In addition, the analysis[52] of available experimental data measured in the most metallized state of hydrogen[53], showed that there is a possibility that this state exhibits strong nonadiabatic effects[41–43]. This implies that the observed $T_c$ in the experiment will be significantly suppressed from the value calculated by theoretical approach where the nonadiabatic effects were not counted.

However, hypothetical possibility, that some alloys with high concentration of hydrogen (named superhydrides) can also exhibit mentioned above properties #1-#3, had been expressed years ago[46,54].

After nearly five decades of experimental quest[48,49] of the *terra incognita* of room temperature superconductivity, Drozdov *et al*[55] experimentally discovered the one in highly compressed $H_3S$. To date, the highly compressed $LaH_{10}$ holds record high $T_c$ (with $T_{c,onset}$ = 280 K[56,57] at pressure $P \sim 200$ GPa) for ever known superconductors[48,58]. While the zero resistance and the Meissner effect in superhydrides had been registered already in the first report on $H_3S$[55], later these physical phenomena have been confirmed in $LaH_{10}$[57,59] and $CeH_9$[60]. The third fundamental phenomenon in the superconductors, which is the flux trap[61–63] effect, recently has discovered in $H_3S$[64], $LaH_{10}$[64] and $CeH_9$[60].

Returning now to the La-H binary system, we need to note that this system has a very rich phase diagram[56,57,65,66]. This was already shown in the first studies by Drozdov *et al*[56]. In the following extended study by Sakata *et al* [65] the multiphase feature of samples in the La-H system has been shown with a great clarity by reporting at least five fundamentally different XRD scans for La-H samples with the onset of transition temperature within a range of $65\ K \leq T_{c,onset} \leq 112\ K$. Later the crystalline structure for seven high-pressure $La_xH_y$ phases have been identified by Laniel *et al*[66],

Recently, Guo et al[67] showed that there is a superhydride phase in La-H binary system, with a stoichiometry of La$_4$H$_{23}$. This phase exhibits A-15 (beta tungsten) arrangement of lanthanum ions[43], and $T_{c,zero}$ = 81 K at pressure $P$ = 118 GPa. Thus, A-15 superconductors' family can be extended by the high-temperature hydride superconductor La$_4$H$_{23}$. Despite the La$_4$H$_{23}$ has much lower $T_c$ in comparison with its near-room-temperature counterpart LaH$_{10}$, the La$_4$H$_{23}$ holds the record high $T_c$ within A-15 family.

Soon after the report by Guo et al[67], Cross et al[68] confirmed the high-temperature superconductivity in highly compressed La$_4$H$_{23}$ phase with measured $T_{c,R\to 0\,\Omega} \cong 60\,K$ ($P$ = 95 GPa), which was defined by strict resistive criterion:

$$\frac{|R(T)-R(T_{c,onset})|}{R(T_{c,onset})} \leq 1 \times 10^{-3} \qquad (3)$$

Recently two research groups[69,70] specialized in the first-principles calculations of high-pressure superconductors showed that the transition temperature of highly compressed superconductors should be affected by the crystalline lattice distortions (caused by either the presence of vacancies[69], either by the anisotropic crystalline strain[70]). Considering that there is also a dependence of the Debye temperature, $\Theta_D$, from the size of the crystals[71], in this study we extracted the crystalline size, $D$, and the nanostrain, ε, in highly pressurized La$_4$H$_{23}$ superconductors from the XRD data reported by two research groups[67,68]. From our view, these parameters are additional characteristics which can enrich our understanding of the near-room-temperature superconductivity in superhydrides.

Thus, in this paper, we analyzed available experimental data measured in La$_4$H$_{23}$ by two research groups[67,68], and we found that this superconducting phase exhibits:

1. Nanograined structure, with average size of coherent-scattering regions, $D$, varied in the range $5.5\,nm \leq D \leq 35\,nm$;

2. Low nanocrystalline strain, $\varepsilon$, which is varied in the range $-0.003 \leq \varepsilon \leq 0.003$ (where negative $\varepsilon$ can be interpreted as the state with high concentrations of hydrogen vacancies);

3. Relatively low Debye temperatures, $\Theta_D(P = 95\ GPa) \cong 500\ K$, and $\Theta_D(P = 118\ GPa) \cong 860\ K$, which implies that the La$_4$H$_{23}$ is strong coupled superconductor with high electron-phonon coupling constant $1.5 \leq \lambda_{e-ph} \leq 2.55$;

4. Moderate level of the nonadiabaticity, $0.18 \leq \frac{\Theta_D}{T_F} \leq 0.22$ (where $T_F$ is the Fermi temperature).

5. Deduced ratio of $0.020 \leq \frac{T_c}{T_F} \leq 0.025$ implies that the La$_4$H$_{23}$ phase falls to unconventional superconductors band in the Uemura plot.

## II. Experimental data sources and data analysis tool

Primarily, we performed our analysis for experimental datasets provided by Cross *et al* [68] as free online experimental data source at the University of Bristol data center [68]. $R(T)$ dataset for report by Drozdov *et al* [56] provided as Data Source for Ref. [56]. Data for Guo *et al* [67] and Sakata *et al* [65] reports were digitized from original plots in the papers [65,67]. Each section describes the models and mathematical routines used for the analysis. The Origin software was used to perform all data fits. List of used designations is given in Table I.

**Table. I.** The list of used designations.

| Designation | Meaning | Equation |
|---|---|---|
| $\Theta_D$ | Debye temperature | 1,8,10-12,14,20 |
| $\Theta_E$ | Einstein temperature | 12 |
| $\lambda_{e-ph}$ | Dimensionless constant of the electron-phonon interaction | 1,14-16,18,19 |
| $\mu^*$ | Dimensionless reduced electron–electron interaction constant (the Coulomb pseudopotential) | 1,14-16 |
| $\theta$ | The Bragg angle | 4,5,6 |
| $\beta_i(\theta)$ | Instrumental breadth of the Bragg peaks in the XRD experiment | 6 |
| $U, V, W$ | Dimensionless parameters of the equation for instrumental broadening of the XRD experiments | 6 |
| $k_s$ | Scherrer constant in the Scherrer and Williamson-Hall equation, usually assigned as 0.9 | 5 |
| $\rho_{sat}$ | Saturated resistivity constant in parallel resistivity model | 8 |
| $\rho_0$ | Residual resistivity, $\rho_0 \equiv \rho(T \to 0\ K)$ | 8 |
| $\gamma$ | Sommerfeld coefficient in the equation for temperature dependent heat capacity | 9,11,12 |
| $\beta$ | the amplitudes of the harmonic phonon contribution in the equation for temperature dependent heat capacity | 9,10 |
| $\delta$ | the amplitudes of the anharmonic phonon contribution in the equation for temperature dependent heat capacity | 9 |
| $\varepsilon$ | Crystalline strain at nanoscale level | 5 |
| $\alpha \equiv \dfrac{2 \times \Delta(0)}{k_B \times T_c}$ | Dimensionless superconducting gap-to-transition temperature ratio | 18,19 |

## III. Results

### 3.1. Size-strain analysis

There are no direct macroscopical techniques which can be applied to study the microstructure (at the submicron level) of the sample in the diamond anvil cell (DAC). However, primary structural parameters of the sample in DAC can be extracted from classical Williamson-Hall (WH) analysis[72] of the X-ray diffraction (XRD) data.

We fitted XRD scans to multiple peaks Lorentz function[73–75] (Figures S1,S2):

$$I(2\theta) = I_{background} + \sum_{k=1}^{N} \frac{2 \times I_k}{\pi} \times \frac{\beta_k}{4 \times (2\theta - 2\theta_{peak,k})^2 + \beta_k^2}, \qquad (4)$$

where $I_k$ is the peak area, $2\theta_{peak,k}$ is the peak position, $\beta_k$ is peak integral breadth, and $I_k$, $2\theta_{peak,k}$, and $\beta_k$ are-free fitting parameters. We manually adjusted the $I_{background}$ level for each panel showed in all figures in this study.

In Figure S1 we showed the fit of the XRD data reported by Cross *et al*[68] to Eq. 4, where we designated by thick red curves all peaks described by Cross *et al*[68] as peaks of the La4H23 phase. One can see (Figure S1) that there are many peaks which were not designated to the La4H23 phase by Cross *et al*[68]. This is another confirmation of the findings by Sakata *et al*[65] and Laniel *et al*[66], that there are several La-H phases which can simultaneously exist in the DAC sample. However, we should stress that as it showed by the first-principles calculations by Guo et al [67], high-temperature superconductivity with $T_c \sim 80$ K is associated exclusively with the La$_4$H$_{23}$ phase at the pressure range of $\sim 100$ GPa.

Derived dataset for the peak breadth, $\beta_i(\theta)$, and peak diffraction angle, $\theta_{peak,i}$, for the La$_4$H$_{23}$ phase was fitted to WH equation[72] (where we assumed that the instrumental broadening, $\beta_i$, is negligible):

$$\beta(\theta, P) = \frac{k_s \times \lambda_{X-ray}}{D(P) \times cos(\theta)} + 4 \times \varepsilon(P) \times tan(\theta), \tag{5}$$

where $k_s$ is the Scherrer constant usually assigned as $0.9$[75–78], $\lambda_{X-ray} = 42.5\ pm$ is the wavelength of the radiation used in Ref.[68], and $\lambda_{X-ray} = 41.24\ pm$ is the wavelength of the radiation used in Ref.[67], and $D(P)$ is the mean size of coherent scattering regions, and the $\varepsilon(P)$ is the nanocrystalline strain.

The reason for the assumption that the instrumental broadening, $\beta_i$, can be omitted in both experiments[67,68] because it is small in comparison with the broadening originated from the sample is based on the following facts:

The assumption that the instrumental broadening, $\beta_i$, can be omitted in both experiments[67,68] because it is small in comparison with the broadening originated from the sample is based on the following facts:

1. In synchrotron experiments (despite there are many experimental approaches and instrumental arrangements[78–81]) typical line broadening for the standardized samples and diffraction angles $2\theta \lesssim 25°$ is $0.0025° \leq \beta_{raw} \leq 0.005°$. This implies that the upper limit for the crystalline size, which can be determined from synchrotron XRD data, is $\sim 2.0\ \mu m$[79] (this is in ~10 times larger than the upper size limit for the laboratory machines data[73–77]).

2. Despite the XRD peaks in the high-pressure synchrotron experiments[78,82–85] are broader in comparison with the ambient pressure experiments[74,75,78–81], it should be noted, that it is difficult to ensure that at high-pressure conditions the sample remains the initial size of coherent scattering regions and the strain. Reported values for the resolution of the high-pressure synchrotron experiments are within the range of $0.01° \lesssim \beta_i \lesssim 0.04°$ for $2\theta \lesssim 25°$.

3. Considering that both research groups[67,68] did not report the instrumental broadening in their experiments, we can estimate this experimental characteristic in the following way. From reported XRD scan[68] we choose three narrowest peaks (Figure S3), which are not the reflections of the La$_4$H$_{23}$ phase (for instance, the peak at $2\theta \cong 7.640°$ has $\beta_{raw} = 0.024°$ (Figure S3)). The fit of this $\beta_{raw}(2\theta)$ dataset to Eq. 5 (Figure 1,a) shows that the crystalline size, associated with these reflections, is $D = 111\ nm$ and the strain is $\varepsilon = 3 \times 10^{-4}$.

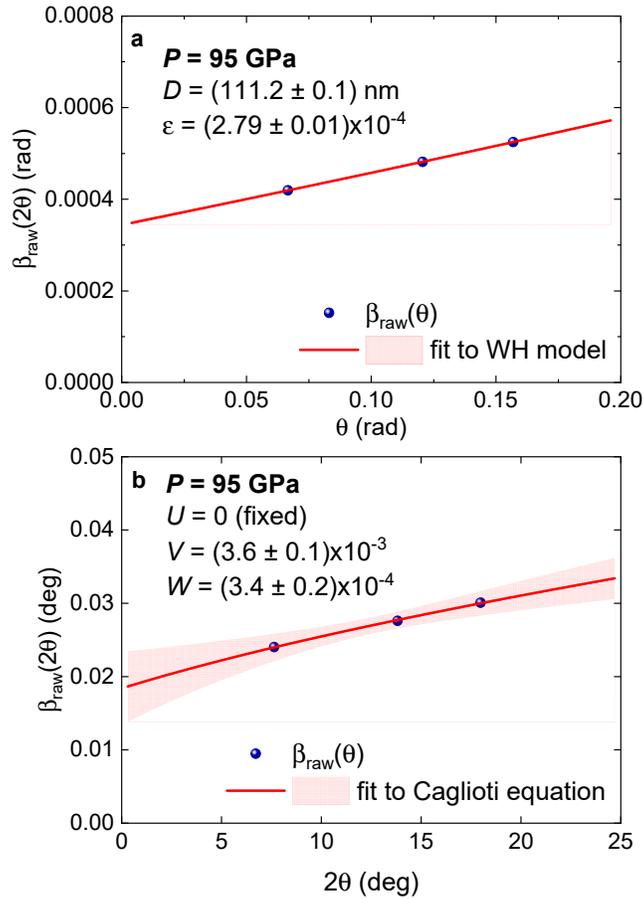

**Figure 1.** XRD peaks breadth, $\beta(\theta)$, for three narrowest peaks recorded by Cross *et al*[68] and the data fit to (a) Williamson-Hall equation[72] (Eq. 5); and (b) to Caglioti equation[86] (Eq. 6). 95% confidence bands are shown by pink areas. Deduced parameters are shown. Fit quality is (a) 0.99999; and (b) 0.9987.

Because it is unlikely, that at harsh conditions of the La$_4$H$_{23}$ phase synthesis in the DAC[67,68], some phase can exhibit conditions for perfect growth of large crystals, the revealed crystalline size $D = 111\ nm$ should not be associated with the upper limit for the experimental resolution. Instead, it is the broadening where some part of it originates from the sample size/strain and another part is from the instrument. However, because we do not have another suitable dataset, we assumed that this dataset represents the upper limit, $\beta_{i,upper}(2\theta)$, for the instrumental broadening for Cross *et al*[68] experiment.

We can use this dataset to estimate parameters in the Caglioti equation[86]:

$$\beta_{i,upper}(2\theta) = \sqrt{U \times tan^2\left(\frac{2\theta}{2}\right) + V \times tan\left(\frac{2\theta}{2}\right) + W} \qquad (6)$$

where $U$, $V$, and $W$ are free-fitting parameters. Both research groups[67,68] conducted experiments for $2\theta \leq 22°$ and, because of the limited number of data points (only three), we omitted the term of $U \times tan^2(2\theta)$. Result of the fit is shown in Figure 1(b), where deduced $V$ and $W$ parameters are shown. In Figure 2(a-c) we showed the extracted $\beta_{raw}(\theta, P = 95\ GPa)$ data for the La$_4$H$_{23}$ phase. In Figure 2 (d-f) we showed corrected peaks breadth by the equation:

$$\beta(\theta, P = 95\ GPa) = \beta_{raw}(2\theta, P = 95\ GPa) - \beta_{i,upper}(2\theta, P = 95\ GPa) \qquad (7)$$

In Figure 2 we fitted data to the Williamson-Hall equation (Eq. 5).

Figure 2 shows that deduced $D(P = 95\ GPa) \sim 30\ nm$, and the strain is low $\varepsilon(P = 95\ GPa) \leq 0.005$. One can see that when fits performed for the condition when both parameters are free (Figure 2(a,d)), then $2\sigma$ uncertainties for both parameters are large. Based on that in Figures 2(b,e) we restricted fits to reveal the minimum size of the nanocrystals, $14\ nm \leq D_{min}(P = 95\ GPa) \leq 17\ nm$ (by applying the condition of $\varepsilon(P = 95\ GPa) \equiv 0$).

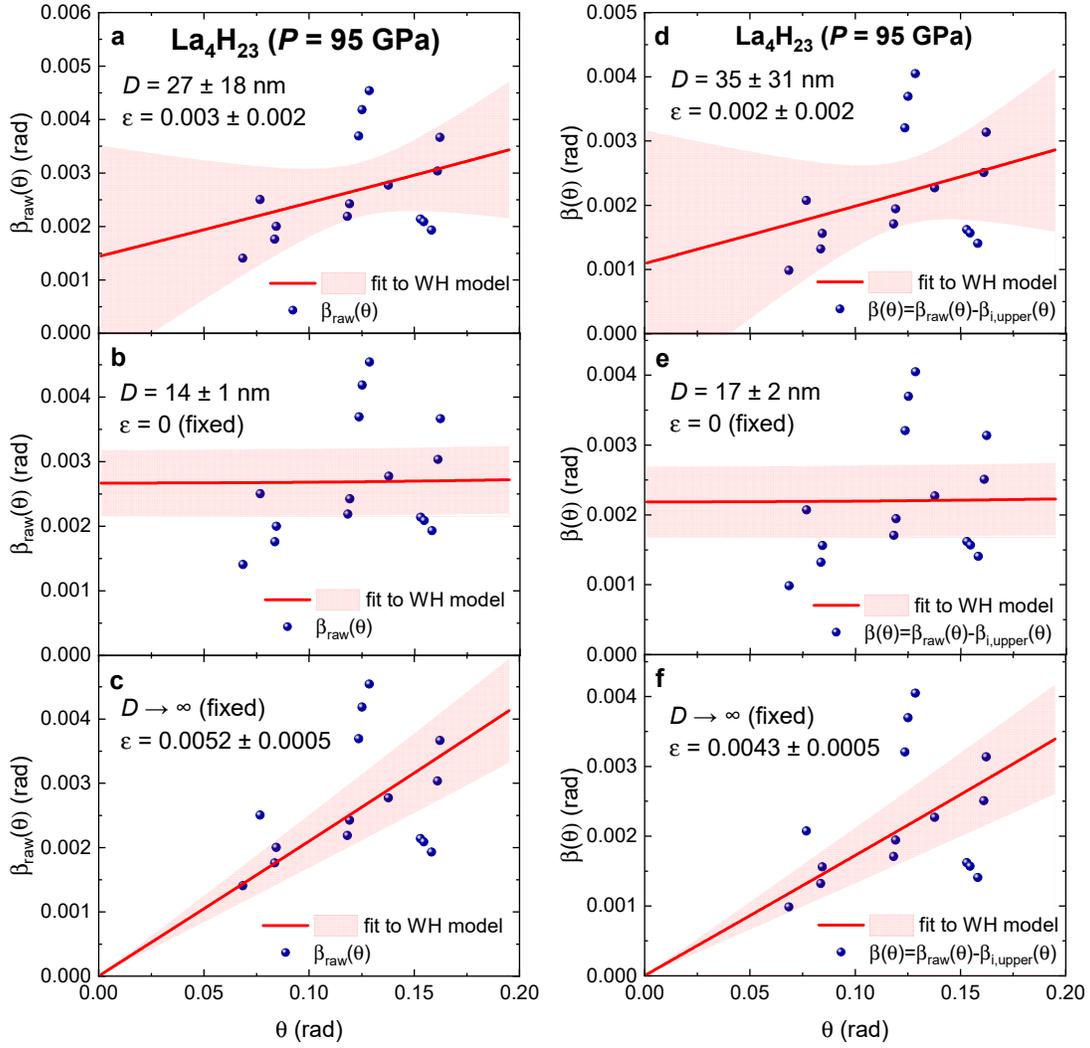

**Figure 2.** XRD peaks breadth (a-c) $\beta_{raw}(\theta)$ and (d-f) $\beta(\theta) = \beta_{raw}(\theta) - \beta_{i,upper}(\theta)$ data and data fits to the Williamson-Hall equation[72] (Eq. 5) for highly compressed single crystal La$_4$H$_{23}$ ($P$ = 95 GPa). Raw XRD scans reported by Cross *et al*[68]. 95% confidence bands are shown by pink areas. (a,d) $\beta(\theta, P)$ data fit to Eq. 5 when $D(P)$ and $\varepsilon(P)$ are free-fitting parameters. (b,e) $\beta(\theta, P)$ data fit to Eq. 3 for the condition $\varepsilon(P) \equiv 0$. (c,f) $\beta(\theta, P)$ data fit to Eq. 5 for the condition $D(P) \to \infty$.

In Figures 2(c,f) we restricted the fit to estimate the maximum of the strain in nanocrystals, $0.004 \leq \varepsilon_{max}(P = 95\ GPa) \leq 0.005$ (by applying the condition of $D(P = 95\ GPa) \to \infty$). In overall (Figure 2), our analysis showed that the nanocrystalline strain in the La$_4$H$_{23}$ phase at $P = 95\ GPa$ is low, because its maximum possible value of $\varepsilon_{max}(P = 95\ GPa) = 0.005 \pm 0.001$ is approximately equal to the value determine by the same WH technique in high-quality epitaxial undoped YBa$_2$Cu$_3$O$_{7-\delta}$ films[87].

In the similar way, we analyzed XRD data reported by Guo *et al*[67]. In Figure 3 we showed the WH analysis for the La$_4$H$_{23}$ (118 $GPa$) sample, where we showed all peaks which we deduced from the experimental scan (Figure S2). Figure 3 shows data fit to Eq. 5, where in panels (a,c) one can see a negative value for the crystalline strain $\varepsilon(118\ GPa) = -0.003 \pm 0.003$.

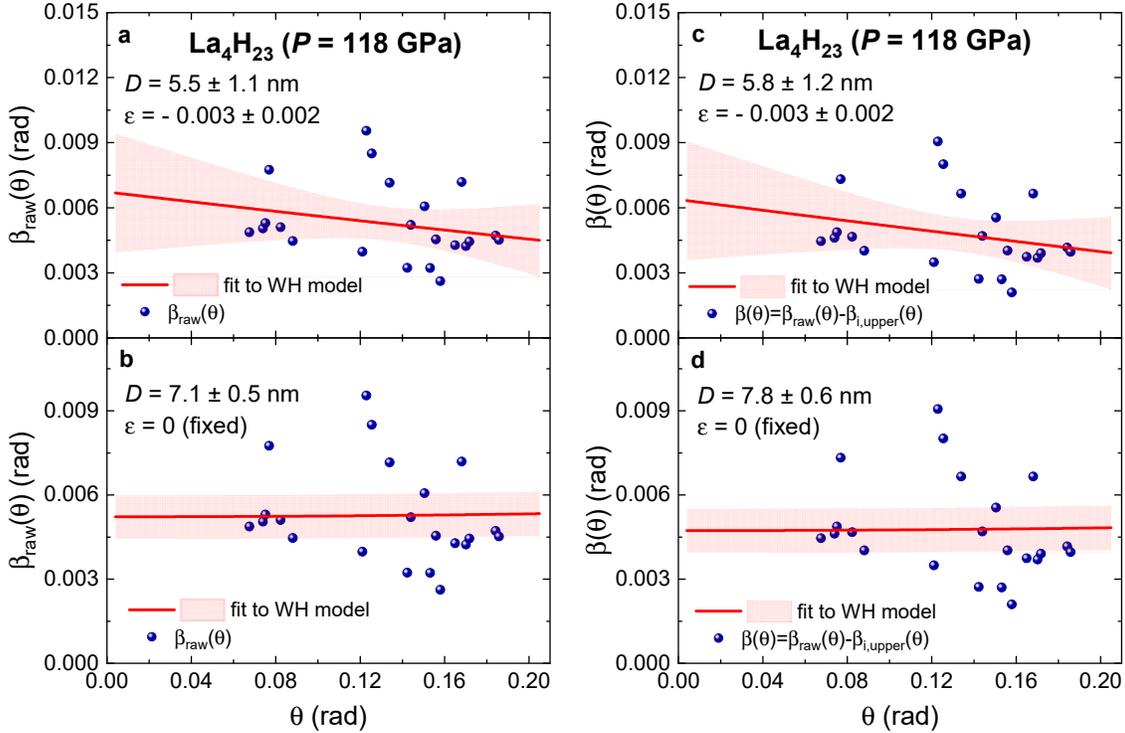

**Figure 3.** XRD peaks breadth (a,b) $\beta_{raw}(\theta)$ and (c,d) $\beta(\theta) = \beta_{raw}(\theta) - \beta_{i,upper}(\theta)$ data and data fits to the Williamson-Hall equation[72] (Eq. 5) for highly compressed single crystal La$_4$H$_{23}$ ($P$ = 118 GPa). Raw XRD scans reported by Guo *et al*[67]. Pink areas show 95% confidence bands. (a,c) $\beta(\theta, P)$ data fit to Eq. 5 when $D(P)$ and $\varepsilon(P)$ are free-fitting parameters. (b,d) $\beta(\theta, P)$ data fit to Eq. 5 for the condition $\varepsilon(P) \equiv 0$.

Negative value for the strain is not what often reported, however, this is not unusual[88], and the $\varepsilon < 0$ values are interpreted as the absence of the strain in the crystal[88]. Considering that the 2σ uncertainty for $\varepsilon$ value is the same as the value itself, we concluded that the sample synthesised and studied by Guo *et al* [67] has very low or no nanostrain.

Figure 3 also shows that the instrumental broadening, $\beta_i(\theta)$, for synchrotron experiments[67], for sample with crystalline size of $D = 6\ nm$ can be completely omitted.

By omitting the stain contribution in the broadening, we can determine the grain size, $D(118\ GPa) = 7\ nm$, with better accuracy by performing the fit at the condition of $\varepsilon \equiv 0$ (Fig. 3 b,d). Considering that the lattice parameter $a(118\ GPa) = 0.614\ nm$, we can conclude that this sample has a nanogranular structure $D(118\ GPa) \cong 11 \times a$.

[68]In overall, we found that synthesized La$_4$H$_{23}$ samples by both research groups have nanograin structure with average grain size in the range of $5.5\ nm \lesssim D \leq 35\ nm$ and low nanocrystalline strain $|\varepsilon| \leq 0.003$.

### 3.2. Debye temperature

As we mentioned above, the Debye temperature, $\Theta_D$, is one of primary parameters in the theory of the electron-phonon mediated superconductivity[38]. *De facto* the standard technique to determine the Debye temperature, $\Theta_D$, for samples in DAC[89–91] is the fit of the normal part of the temperature dependent resistance, $R(T)$, to the saturated resistance model[92–94] (where the $\Theta_D$ is a free-fitting parameter):

$$\rho(T) = \frac{1}{\frac{1}{\rho_{sat}} + \frac{1}{\rho_{Debye}(T)}} = \frac{1}{\frac{1}{\rho_{sat}} + \frac{1}{\left(\rho_0 + A \times \left(\frac{T}{\Theta_D}\right)^5 \times \int_0^{\frac{\Theta_D}{T}} \frac{x^5}{(e^x-1)(1-e^{-x})}dx\right)}} \quad (8)$$

where $\rho_{sat}$, $\rho_0$, $A$, and $T_c$ are free-fitting parameters.

Recently Watanabe *et al* [95] reported a good agreement between the $\Theta_D$ deduced from the fit of the $R(T)$ data to the Eq. 8 and from the fit of the low-temperature normal state specific heat capacity, $C_p(T)$, in the η-carbide-type oxide Zr$_4$Pd$_2$O. In the latter technique, the $C_p(T)$ is fitted to the equation:

$$C_p(T) = \gamma \times T + \beta \times T^3 + \delta \times T^5 \quad (9)$$

where $\gamma$ is the Sommerfeld coefficient, $\beta$ and $\delta$ are the amplitudes of the phonon contributions for the harmonic and anharmonic terms, respectively; and the $\Theta_D$ is calculated by the equation:

$$\Theta_D = \left(\frac{12\pi^4 N R_{gc}}{5\beta}\right)^{\frac{1}{3}} \tag{10}$$

where $N$ is the number of atoms per formula unit and $R_{gc} = 8.31\ JK^{-1}mol^{-1}$ is the universal gas constant. It is important to noted that the approach expressed by Eqs. 9-10 is the standard way to determine the Debye temperature from the total specific heat data [95–102].

Advanced approach [103–105] is to fit the $C_p(T)$ data to the Debye equation:

$$C_p(T) = \gamma \times T + 9 \times R_{gc} \times N \times \left(\frac{T}{\Theta_D}\right)^3 \int_0^{\frac{\Theta_D}{T}} \frac{x^4 e^x}{(e^x-1)^2} dx \tag{11}$$

where all parameters defined above, or to multichannel Debye-Einstein equation:

$$C_p(T) = \gamma \times T + 9 \times R_{gc} \times \sum_{i=1}^{M} A_i \left(\frac{T}{\Theta_{D,i}}\right)^3 \int_0^{\frac{\Theta_{D,i}}{T}} \frac{x^4 e^x}{(e^x-1)^2} dx + 3 \times R_{gc} \times \sum_{j=1}^{P} B_j \left(\frac{\Theta_{E,j}}{T}\right)^2 \frac{e^{\left(\frac{\Theta_{E,j}}{T}\right)}}{\left(e^{\left(\frac{\Theta_{E,j}}{T}\right)}-1\right)^2} \tag{12}$$

where $A_i$ and $B_j$ are constants (depended from given crystalline structure and chemical composition), $M$ and $P$ are number of the channels for the Debye modes and the Einstein modes, respectively; $\Theta_{D,i}$ is the Debye temperature of the $i$-channel, $\Theta_{E,j}$ is the Einstein temperature of the $j$-channel. The use of the Eq. 11 requires high sensitivity measurements of the $C_p(T)$ and an addition, all measurements should be performed in a wide temperature range, $0 < T \lesssim \Theta_E$, with a small temperature step, $\Delta T \sim \frac{\Theta_D}{500}$, between measurements, at least at low-$T$ region of normal state [106].

Because DAC has significantly larger thermal mass in comparison with the mass of the sample, this is practically impossible to extract the contribution of the sample in the total $C_p(T)$ from experimental measurements of the total heat capacity. Thus, to extract the Debye temperature for samples in DAC, temperature dependent resistive measurements are fitted to the Eq. 8.

To answer a possible question about the comparison of the deduced $\Theta_D$ values extracted from the fit of the $C_p(T)$ data to Eq. 11 and from the fit of the $\rho(T)$ data to Eq. 8, in Figure 4 we showed the $C_p(T)$ and $\rho(T)$ data fits for the cubic centrosymmetric η-carbide $Nb_4Rh_2C_{1-\delta}$ (raw datasets were reported by Ma *et al*[107]). Deduced $\Theta_D = (290 \pm 2)\,K$ (from $C_p(T)$ data) and $\Theta_D = (312 \pm 3)\,K$ (from $\rho(T)$ data) are evidences that both methods (i.e. Eqs. 8,11) can be used to extract the Debye temperature from experimental data. For clarity, in Figure 4,c we showed two components of the $\rho(T)$ data fit, $\rho_{sat}$ and $\rho_{Debye}(T)$.

In Figure 5 we showed the $R(T, P = 98\,GPa)$ curves[68] for compressed $La_4Hi_{23}$ and data fits to Eq. 8. Derived Debye temperature, $445\,K \leq \Theta_D \leq 583\,K$, is significantly lower than the values of $1310\,K \leq \Theta_D \leq 1675\,K$ determined for the $LaH_{10}$ phase, which has $T_c > 200\,K$[91].

Surprisingly enough, we found that the sample designated by Drozdov *et al*[56] as $LaH_{x>3}$ (compressed at pressure $P = 150\,GPa$ and it has a designation of Sample 11[56]) exhibits very close Debye temperature $\Theta_D = 655 \pm 2\,K$ and $T_{c,0.05} = 66.2\,K$ (Figure 6). This dataset was also analysed in Ref.[91].

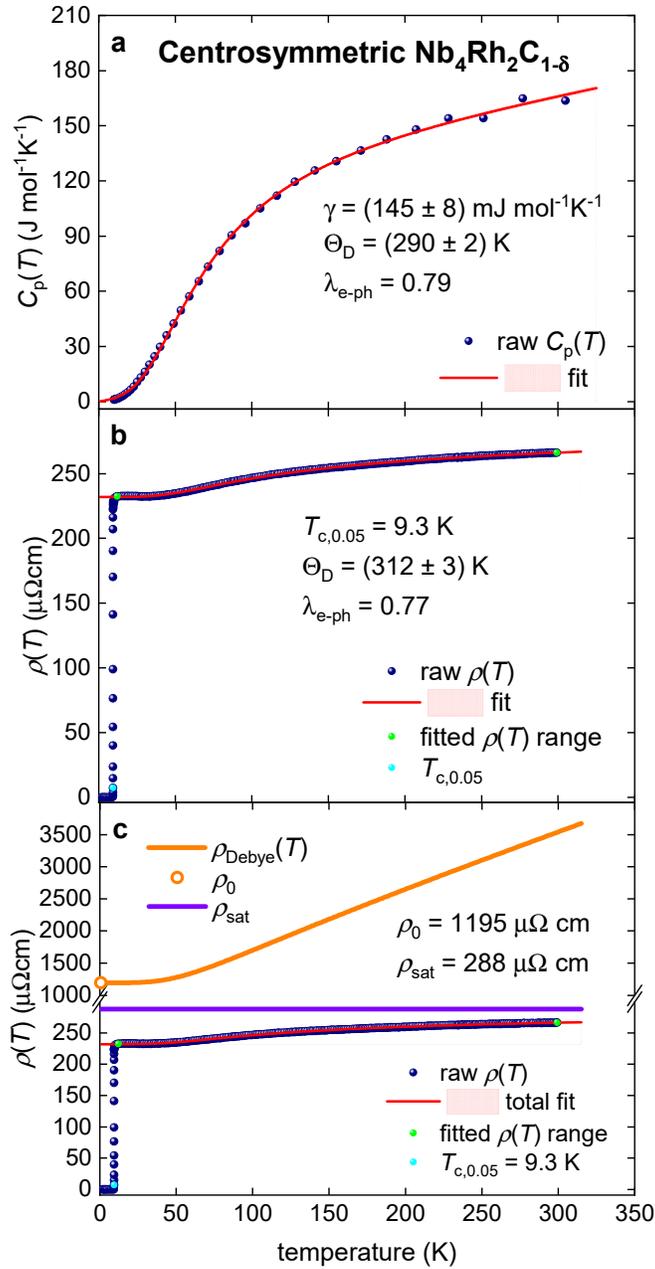

**Figure 4.** (a) Temperature dependent heat capacity, $C_p(T)$, and data fit to Eq. 11 for cubic centrosymmetric η-carbide $Nb_4Rh_2C_{1-\delta}$ (raw data reported by Ma *et al*[107]). (b) Temperature dependent resistivity, $\rho(T)$, and data fit to Eq. 8 for $Nb_4Rh_2C_{1-\delta}$ (raw data reported by Ma *et al*[107]). (c) Temperature dependent resistivity, $\rho(T)$, data fit to Eq. 8, and two components of this fit ($\rho_{sat}$ and $\rho_{Debye}(T)$) for $Nb_4Rh_2C_{1-\delta}$ (raw data reported by Ma *et al*[107]). 95% confidence bands are shown by pink areas. Green balls indicate the bounds for which $\rho(T)$ data was used for the fit. Cyan balls indicate $T_{c,0.05}$. Fit quality for all panels is better than 0.9995.

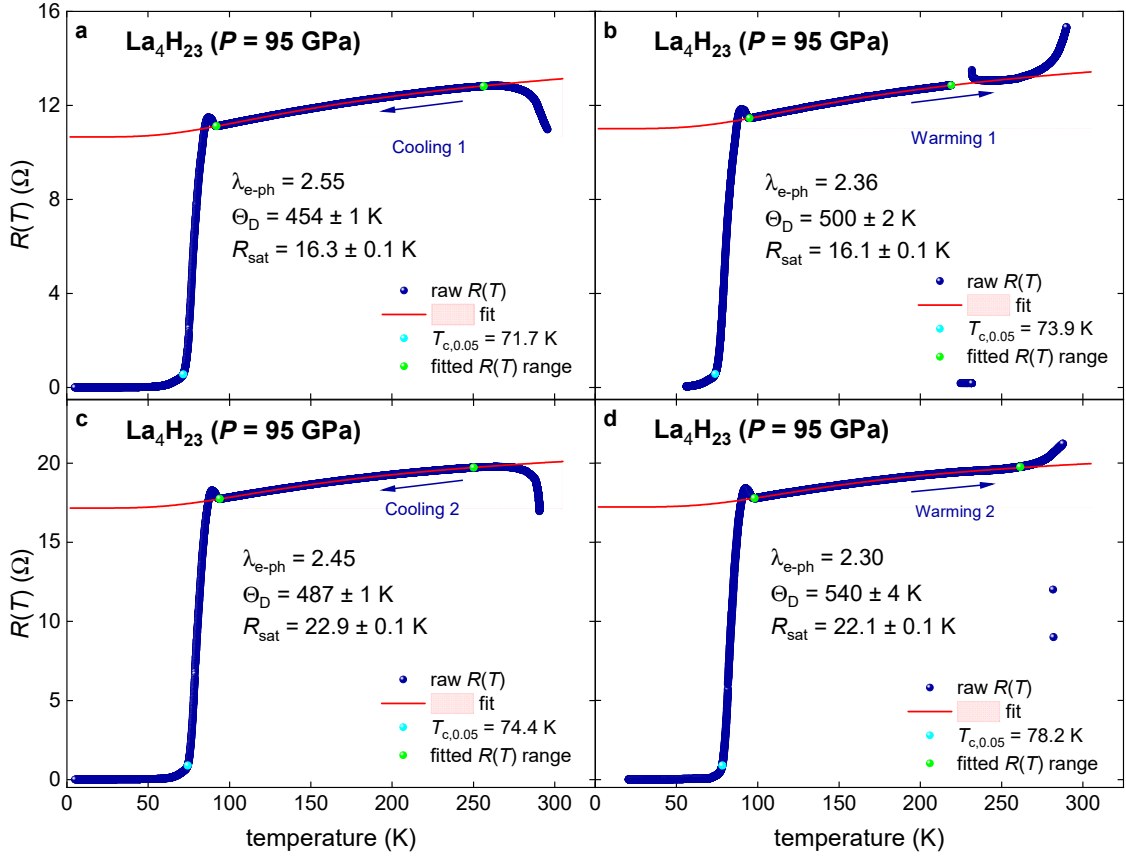

**Figure 5.** Temperature dependent resistance, $R(T,P=98\ GPa)$, measured in compressed La$_4$Hi$_{23}$ and data fits to Eq. 8 (raw data reported by Cross *et al*[68]). 95% confidence bands are shown by pink areas. Green balls indicate the bounds for which $R(T)$ data was used for the fit. Green balls indicate fitted data range. Cyan balls indicate $T_{c,0.05}$. Fit quality for all panels is better or equal to 0.9995. (a) – cooling 1; (b) – warming 1; (c) – cooling 2; (b) – warming 2. Derived $\lambda_{e-ph}$ are for $\mu^* = 0.13$.

Similar surprise is the extracted Debye temperature $\Theta_D = 517 \pm 13\ K$ for the $\rho(T)$ curve reported by Sakata *et al*[65] for LaH$_{x<10}$ (compressed at pressure $P = 170\ GPa$) (Figure 7). The deduced value is practically undistinguishable from the $\Theta_D$ values deduced for four samples of the La$_4$H$_{23}$ phase synthesized by Cross *et al*[68] (Figure 5).

The fit of the reported $R(T)$ data measured in the La$_4$H$_{23}$ ($114\ GPa$) (data reported by Guo *et al*[67]) is shown in Figure 8. Derived Debye temperature is $\Theta_D(114\ GPa) = 904 \pm 69\ K$.

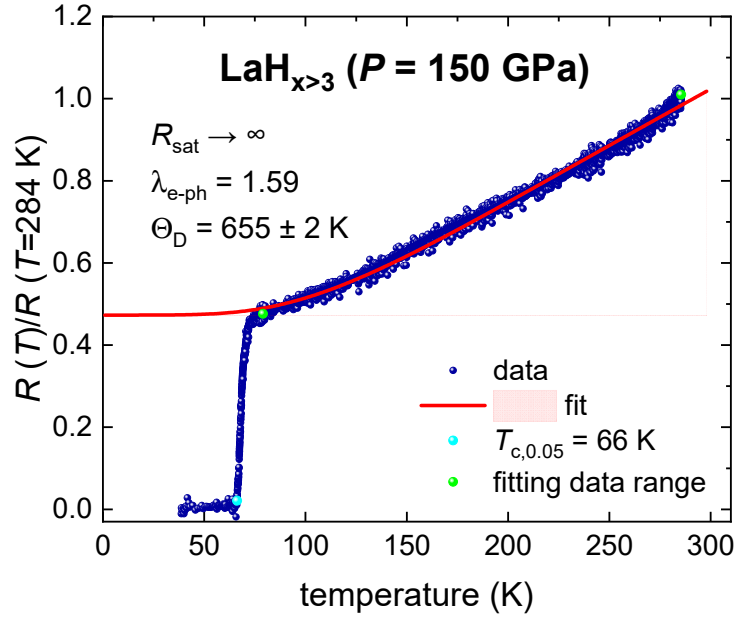

**Figure 6.** Temperature dependent resistance, $R(T,P=150$ GPa), measured in compressed $LaH_x$ (x > 3) sample and data fit to Eq. 8 (raw data reported by Drozdov *et al*[56]; this sample was designated as Sample 11 by Drozdov *et al*[56]). Pink area shows 95% confidence band. Green balls indicate fitted data range. Cyan balls indicate $T_{c,0.05}$. Fit quality is 0.9955. Derived $\lambda_{e-ph}$ is for $\mu^* = 0.13$. See also analysis of the same data in Ref.[91].

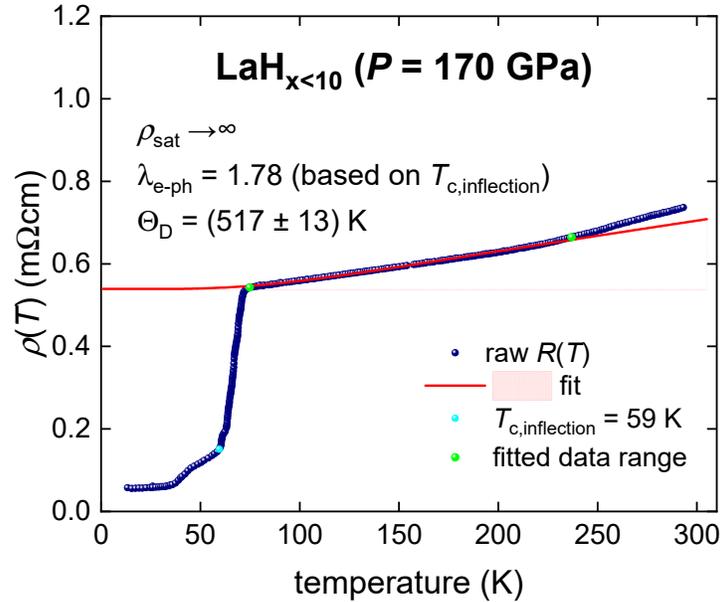

**Figure 7.** Temperature dependent resistance, $\rho(T,P=170$ GPa), measured in compressed $LaH_x$ (x < 10) sample and data fit to Eq. 8 (raw data reported by Sakata *et al*[65] in their Figure 3). Pink area shows 95% confidence band. Green balls indicate fitted data range. Cyan balls indicate $T_{c,onset} = 75\ K$. Fit quality is 0.9971. Derived $\lambda_{e-ph}$ is for $\mu^* = 0.13$.

Based on results showed in Figures 5-8 we can propose that it is quite possible that the Sample 11 reported by Drozdov *et al*[56] and the sample by Sakata *et al* [65] are *de facto* first synthesized and studied samples of the La$_4$Hi$_{23}$ phase in the literature.

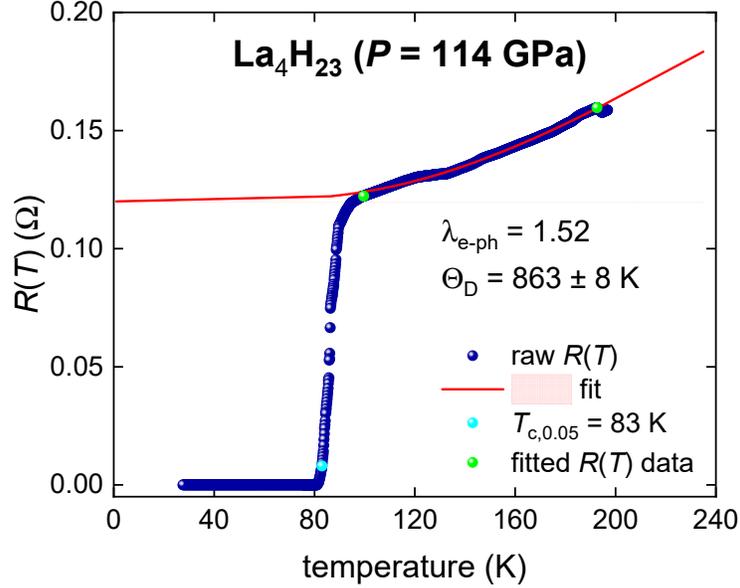

**Figure 8.** Temperature dependent resistance, $R(T,P=114$ GPa), measured in compressed La$_4$Hi$_{23}$ and data fits to Eq. 8 (raw data reported by Guo *et al*[67]). 95% confidence band is shown by pink area. Green balls indicate fitted data range. Cyan ball indicates $T_{c,0.05}$. Fit quality is 0.9964. Derived $\lambda_{e-ph}$ is for $\mu^* = 0.13$.

### 3.3. The electron phonon coupling constant

From deduced $\Theta_D$ and $T_{c,0.05}$, which we defined by the criterion:

$$\frac{|R(T)-R(T_{c,onset})|}{R(T_{c,onset})} = 0.05, \qquad (13)$$

the electron-phonon coupling constant, $\lambda_{e-ph}$, can be determined as the root of advanced McMillan equation[91]:

$$T_c = \left(\frac{1}{1.45}\right) \times \Theta_D \times e^{-\left(\frac{1.04(1+\lambda_{e-ph})}{\lambda_{e-ph}-\mu^*(1+0.62\lambda_{e-ph})}\right)} \times f_1 \times f_2^*, \qquad (14)$$

where

$$f_1 = \left(1 + \left(\frac{\lambda_{e-ph}}{2.46(1+3.8\mu^*)}\right)^{3/2}\right)^{1/3}, \qquad (15)$$

$$f_2^* = 1 + (0.0241 - 0.0735 \times \mu^*) \times \lambda_{e-ph}^2, \qquad (16)$$

where $\mu^*$ is the Coulomb pseudopotential. In this work, we fixed $\mu^* \equiv 0.13$, because this is a typical value for highly compressed electron-phonon mediated superconductors[108–110]. (It can be mentioned that Eqs. 14-16 are more complicated in comparison with Eq. 1, however, these equations remain primary dependences $T_c \propto \Theta_D \times e^{-\frac{1}{\lambda_{e-ph}}}$).

Derived $\lambda_{e-ph}$ values for the LaH$_x$ samples are shown in Figs. 4-8, where in Figure 7, the transition temperature was assumed to be at the inflection point of the $R(T)$ curve.

### 3.4. The ground state coherence length

To deduced the ground state coherence length, $\xi(0)$, which is one of two fundamental lengths in any superconductor, we fitted the data for upper critical field, $B_{c2}(T)$, to the equation proposed by Baumgartner et al[111]:

$$B_{c2}(T) = \frac{1}{0.693} \times \frac{\phi_0}{2\pi\xi^2(0)} \times \left(\left(1-\frac{T}{T_c}\right) - 0.153 \times \left(1-\frac{T}{T_c}\right)^2 - 0.152 \times \left(1-\frac{T}{T_c}\right)^4\right) \quad (17)$$

Where $\phi_0 = \frac{h}{2e}$ is the superconducting flux quantum, $h = 6.626 \times 10^{-34} \, J \cdot s$ is Planck constant, $e = 1.602 \times 10^{-19} \, C$, and $\xi(0)$, and $T_c \equiv T_{c,0.05}(B=0)$ are free fitting parameters.

We extracted raw $B_{c2}(T)$ datasets from the $R(T,B)$ datasets by the criterion described by Eq. 13. In Fig. 9 we showed the $B_{c2}(T)$ data and data fits to Eq. 17, from which the derived $(2.1 \pm 0.1) \, nm \leq \xi(0) \leq (3.0 \pm 0.1) \, nm$. [68]

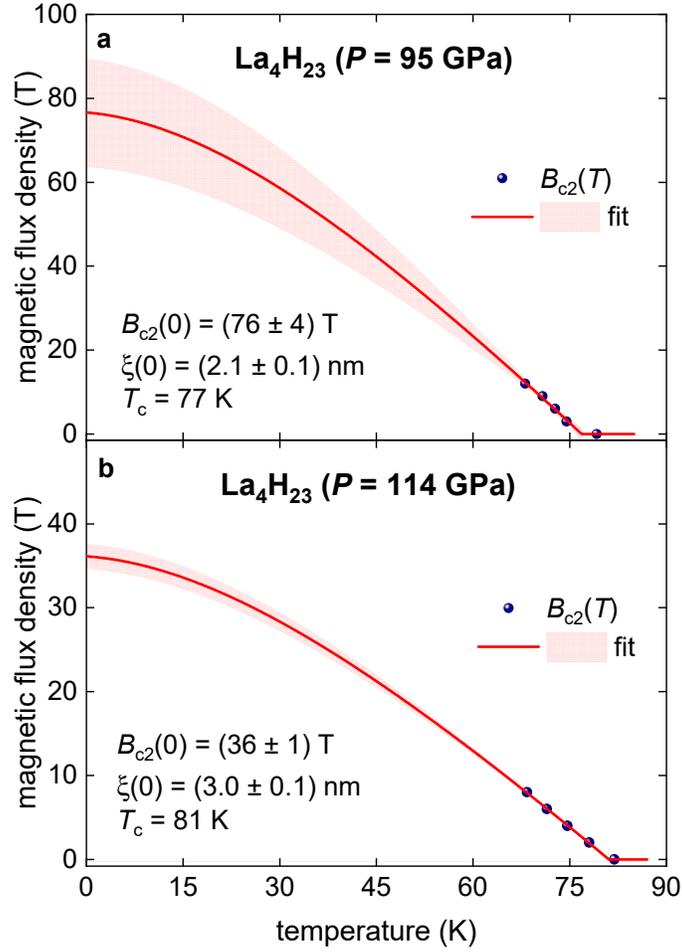

**Figure 9.** The upper critical field data, $B_{c2}(T)$, for compressed La$_4$H$_{23}$ and data fits to Eq. 17. Raw $R(T)$ data reported by (a) Cross et al[68] and (b) Guo et al[67]. Pink area shows 95% confidence band.

### 3.5. The Fermi temperature

To calculate the Fermi temperature, we used the equation[112,113]:

$$T_F = \frac{\pi^2 m_e}{8 \cdot k_B} \times (1 + \lambda_{e-ph}) \times \xi^2(0) \times \left(\frac{\alpha \times k_B \times T_c}{\hbar}\right)^2, \tag{18}$$

where $m_e = 9.109 \times 10^{-31}\ kg$ is bare electron mass, $\hbar = 1.055 \times 10^{-34}\ J \cdot s$ is reduced Planck constant, $k_B = 1.381 \times 10^{-23}\ m^2 \cdot kg \cdot s^{-2} \cdot K^{-1}$ is Boltzmann constant, $\alpha \equiv \frac{2 \times \Delta(0)}{k_B \times T_c}$ is the gap-to-transition temperature ratio, where $\Delta(0)$ is the ground state amplitude of the superconducting gap.

In Sections 3.2.-3.4. we determined all terms in the Eq. 18, except the $\alpha \equiv \frac{2 \times \Delta(0)}{k_B \times T_c}$ value. The linear empirical relation proposed in Ref.[114] can estimate this value:

$$\alpha \equiv \frac{2\Delta(0)}{k_B \cdot T_c} = 3.26 + 0.74 \times \lambda_{e-ph} \quad (19)$$

Considering that the free-fitting value of $T_{c,0.05}(95\ GPa) = 77\ K$ in Figure 9,a is close to the observed $T_{c,0.05}(95\ GPa) = 78.2 \pm 0.1\ K$ value in Figure 5,d, we substituted the derived $\lambda_{e-ph} = 2.30$ for this sample (i.e. Fig. 5,d) in the Eq. 19, and calculated $\alpha_{La_4H_{23}} = 4.96$. In the result, we estimated the $T_F(95\ GPa) = 3.06 \times 10^3\ K$ in the La$_4$H$_{23}$ ($P = 95\ GPa$). It should be noted that our calculated the Fermi velocity, $v_F(95\ GPa) = 1.68 \times 10^5\ \frac{m}{s}$, is in remarkable agreement with the value $v_F(95\ GPa) = 1.80 \times 10^5\ \frac{m}{s}$ reported by Cross *et al*[68].

For sample[67] $T_{c,0.05}(114\ GPa) = 81\ K$ (Figure 9,b), $\lambda_{e-ph}(114\ GPa) = 1.52$ (Figure 8), from which calculated $\alpha_{La_4H_{23}} = 4.38$, and $T_F(114\ GPa) = 3.99 \times 10^3\ K$.

### 3.6. Identification plots

Utilizing deduced parameters, we found that the La$_4$H$_{23}$ phase falls to the unconventional superconductors band in the Uemura plot (Figure 10), and it locates near cuprates and another superhydride LaBeH$_8$ ($P = 120$ GPa).

We also deduced the ratios of $\frac{\Theta_D}{T_F}(95\ GPa) = 0.176$ and $\frac{\Theta_D}{T_F}(114\ GPa) = 0.216$ for the La$_4$H$_{23}$ phase. This value is used to locate this phase in the $\frac{\Theta_D}{T_F}$ vs $\lambda_{e-ph}$ diagram (Figure 11) (this type of diagram proposed by Pietronero *et al* [42,43,115,116]).

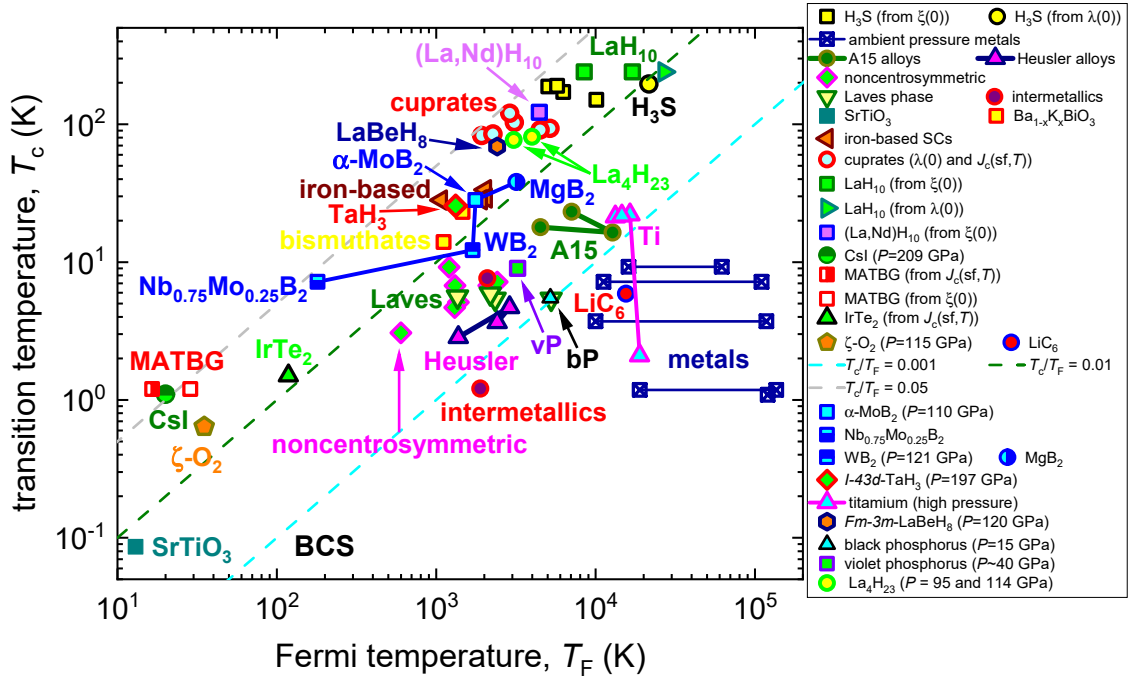

**Figure 10.** Uemura plot where the highly compressed $La_4H_{23}$ phase is shown together with the main families of superconductors: metals, iron-based superconductors, diborides, cuprates, Laves phases, and near-room-temperature superconducting. References to original data can be found in Refs. [43,52,89,112,113,115,117–124].

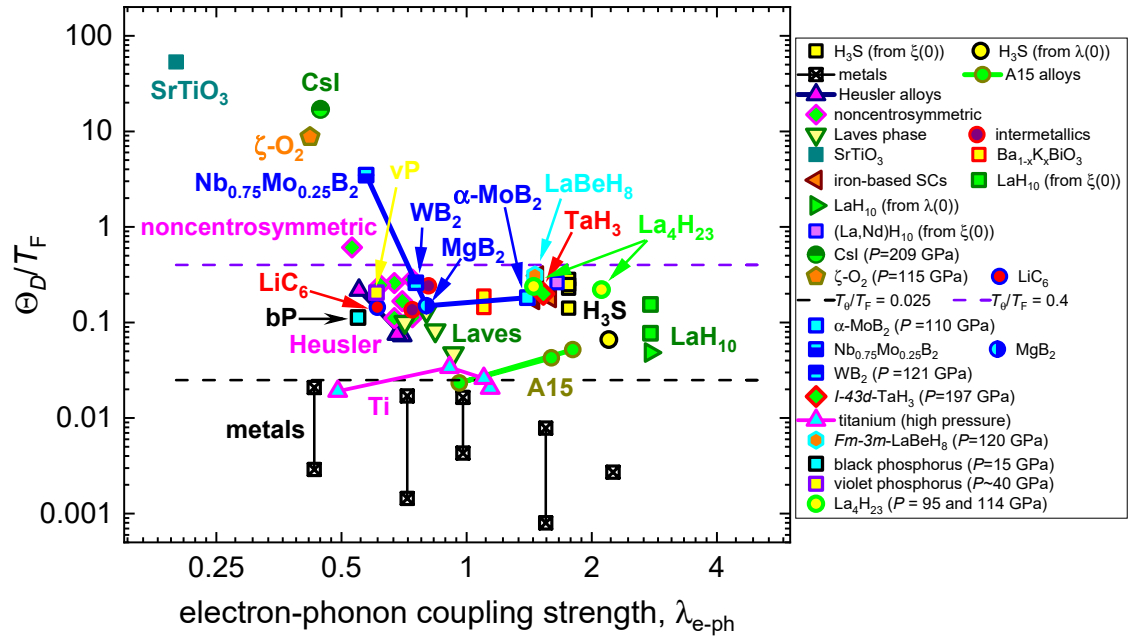

**Figure 11.** The $\frac{\Theta_D}{T_F}$ vs $\lambda_{e-ph}$ plot (this type of plot was proposed in Ref. [42,43,115,116]) where several families of superconductors and the $La_4H_{23}$ phase. References are given in [43,52,89,112,113,115,117–124].

In addition, the derived ratios of $\frac{\Theta_D}{T_F}$ are used to locate the La$_4$H$_{23}$ phase in the $\frac{\Theta_D}{T_F}$ vs $T_c$ diagram (Figure 12) (this type of diagram proposed in Ref. [117,121]). The advantage of this type of plot is that it links three primary thermodynamic quantities in superconductors, which are average energies per particles in the superconductor: the Cooper pairs, electrons, and atomic ions, while the Uemura plot (Figure 10) and the Pietronero plot (Figure 11) link only two characteristic energies per particles in superconductors. All values (and references, for each value) showed in Figures 10-12 are given in Refs. [117,119,121].

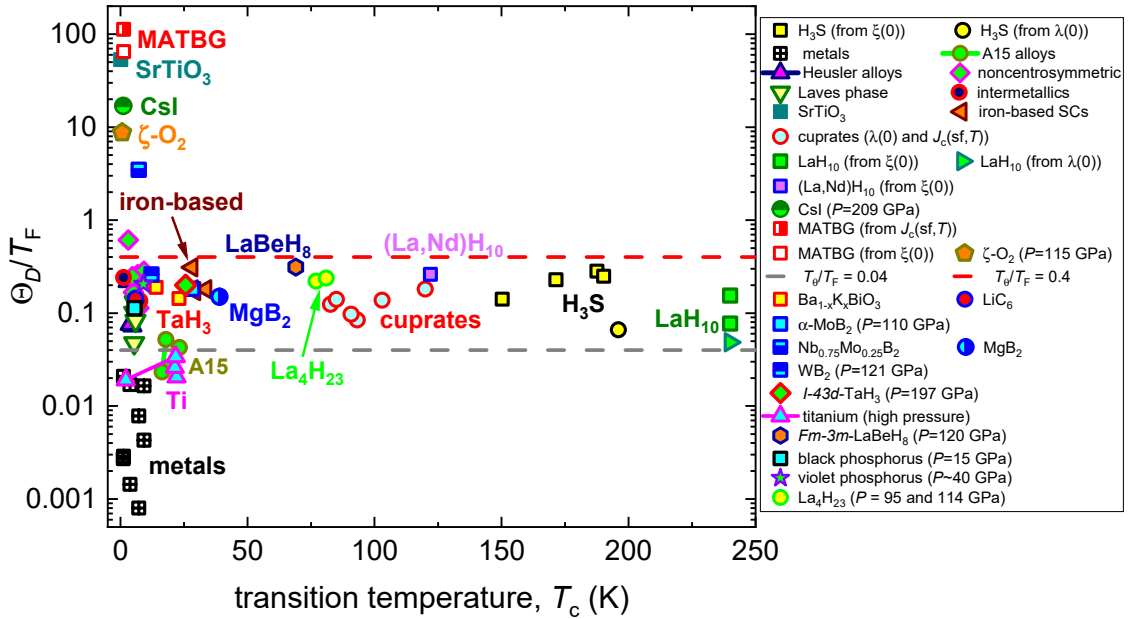

**Figure 12.** The $\frac{\Theta_D}{T_F}$ vs $T_c$ plot (this type of plot was proposed in Ref. [117,121]) for several families of superconductors and highly compressed $La_4H_{23}$. References are given in [43,52,89,112,113,115,117–124].

## IV. Discussion

One can see in Figure 12, that all superconductors with $T_c \geq 20\ K$ exhibit the $\frac{\Theta_D}{T_F}$ ratio within a narrow range, $0.04 \leq \frac{\Theta_D}{T_F} \leq 0.4$. This is perhaps the most interesting finding of this study, because the La$_4$H$_{23}$ phase falls into this narrow band. While Pietronero and co-workers [41–43,115,116,125] prefer to use the value of $\lambda_{e-ph} \times \frac{\Theta_D}{T_F}$, as it showed in their Figure 11 [43], the use of the $\frac{\Theta_D}{T_F}$ ratio, without additional multiplication term $\lambda_{e-ph}$, is also useful, because it utilizes only three fundamental temperatures of any conductor, and it does not use any multiplicative factor associated with any pairing mechanism (for instance, the electron-phonon coupling strength, for the electron-phonon mediated mechanism).

Despite there is no theoretical explanation for this (Figure 12) new empirical fact [117,121], the primary physics behind this observation is more likely related to the issue that high superconducting transition temperature, $T_c$, emerges in materials where the highest energy of the charge carriers exceeds, but not overwhelming, the energy of the coherent oscillations of the crystalline lattice. At this condition (if even in some materials the pairing of the Cooper pairs originates from different from the electron-phonon mechanism, for instance, in cuprates), the Cooper pairs are not disturbed by either very energetic coherent lattice oscillations (in cases of $T_F \ll \Theta_D$), or by uncorrelated random thermodynamic local distortions/fluctuations, which exist at high temperature in materials with low Debye temperature, $\Theta_D \ll T_F$.

Based on this picture, we can describe the physics behind the Uemura plot [122,123,126], as an alternative presentation of the same physical picture, where however, the third fundamental temperature of any superconductor, i.e. the Debye temperature, is missed.

Empirical finding [89,90,112,113,117,118,121,127–129] that all highly compressed superhydrides fall to unconventional superconductors band in the Uemura plot (despite all these superconductors exhibit the electron-phonon pairing mechanism) and which was interpreted [89,90,112,113,117,118,121,127–129] as the evidence for the unconventional pairing mechanism in the superhydrides, in fact reveals the issue that the $\frac{\Theta_D}{T_F}$ ratio for all high-temperature superconductors, $T_c \geq 20\ K$, falls to a narrow range of $0.04 \leq \frac{\Theta_D}{T_F} \leq 0.4$.

Following recent theoretical understanding of the physics in superhydrides and other high-temperature superconductors[43] we can characterize all superhydrides by:

1. highest Debye phonon frequency, $\omega_D \geq 50\ THz$, among all superconducting materials;
2. high electron phonon coupling constant, $\lambda_{e-ph} \geq 1.5$;
3. reasonably high, but not record high, Fermi temperature, $2{,}000\ K \leq T_F \leq 30{,}000\ K$;
4. significant quantum lattice fluctuations [130–133], due to high frequency for the lattice oscillations associated with the lightest chemical element;
5. the presence of a Van Hove singularity close to the Fermi level;
6. moderate level of the nonadiabaticity, $0.05 \leq \lambda_{e-ph} \times \frac{T_D}{T_F} \leq 1.0$;
7. reasonably high $\frac{T_c}{T_F}$ ratio, $0.005 \leq \frac{T_c}{T_F} \leq 0.04$, which implies that in the Uemura plot all superhydrides fall to unconventional superconductors band.

As one can see from #1-#7 above, all electron-phonon mediated hydride superconductors have many properties which are similar or identical to cuprates and pnictides. Thus, from this point of view, all superhydrides can be classified as unconventional superconductors [43,90,112,113,118,128], despite a fact that these superconductors exhibit electron-phonon pairing mechanism (which is clearly demonstrated by prominent isotope effect in direct experiments [55,56]).

This paradoxical understanding is clearly expressed and explained by Cappelluti *et al*[43] now, while Luciano Pietronero presented this theoretical concept at the conference in May 2017[134]. In peer-review form[112,113] (see also[135]) and independently this paradox was first reported as empirical finding derived from the experimental data analysis for the temperature dependent upper critical field in highly compressed $H_3S$ and $LaH_{10}$.

From our point of view, all mentioned above properties for high- and near-room-temperature superconductors (including cuprates, pnictides, diborides, and hydrides) can be summarized in the empirical finding condition, where three fundamental temperatures of any superconductor are obeying the strict condition[117,121]:

$$T_c \geq 20\,K \implies 0.04 \leq \frac{\Theta_D}{T_F} \leq 0.4 \qquad (20)$$

Eq. 20 represents a problem which needs to be explained (graphical representation of Eq. 20 is given in Figure 12). It should be also noted that this understanding/explanation might lie beyond the conventional first-principles calculation studies which is the dominant theoretical approach utilized in modern high-pressure superconductivity [50,67,108,130,136–161] and global view on high-pressure superconductivity, where will be presented hydrides[48–50,55,56,59,60,64–67,70,115,118,127,129–134,136,138–144,147,149,151,152,154–158,160–207], other high-pressure superconductors[90,119,120,208–236] and ambient pressure superconductors[62,95,99,237–273], including amorphous[274–276] and quasicrystals[277–282] would be considered from the unified theoretical concept.

**V. Conclusions**

In this work, we analyzed experimental data reported for highly compressed high-temperature superconducting $La_4H_{23}$ phase. This phase is a new A-15-type phase which simultaneously extends the family of highly pressurized hydrides, the family of A-15 superconductors, and the family of high-pressure superconductors.

We found a good agreement between derived Debye temperature $\Theta_D$ for the $La_4H_{23}$ phase and the $\Theta_D$ deduced for two samples with similar values of $T_c \sim 70$ K a phase reported by Drozdov *et al* [56] as the unknown phase with an approximate stoichiometry of $LaH_x$ (x>3) and by Sakata *et al* [65], who designated the sample stoichiometry as the $LaH_x$ (x<10). From this, we proposed that *de facto* the La$_4$H$_{23}$ phase was first discovered in the experiment by Drozdov *et al* [56] and by Sakata *et al* [65].

We also found that the La$_4$H$_{23}$ phase synthesised and studied by both research groups [67,68] has nanoscale grain size and is very low, or even the absence, of the crystalline strain.

In addition, we also deduced the values:

(1) $0.020 \leq \frac{T_C}{T_F} \leq 0.025$;

(2) $0.18 \leq \frac{\Theta_D}{T_F} \leq 0.22$;

(3) $1.5 \leq \lambda_{e-ph} \leq 2.55$.

**Data availability statement**

The data that supports the findings of this study are available from the corresponding author upon reasonable request.

**Declaration of interests**

The authors declare that they have no known competing financial interests or personal relationships that could have appeared to influence the work reported in this paper.

## Supplementary Materials

## La$_4$H$_{23}$ superconductor: Nanograined structure, low nanocrystalline strain, strong electron-phonon interaction, and moderate level of nonadiabaticity

**I.  XRD scans and data fit to multiple Lorenz function for data reported by Cross *et al*[68].**

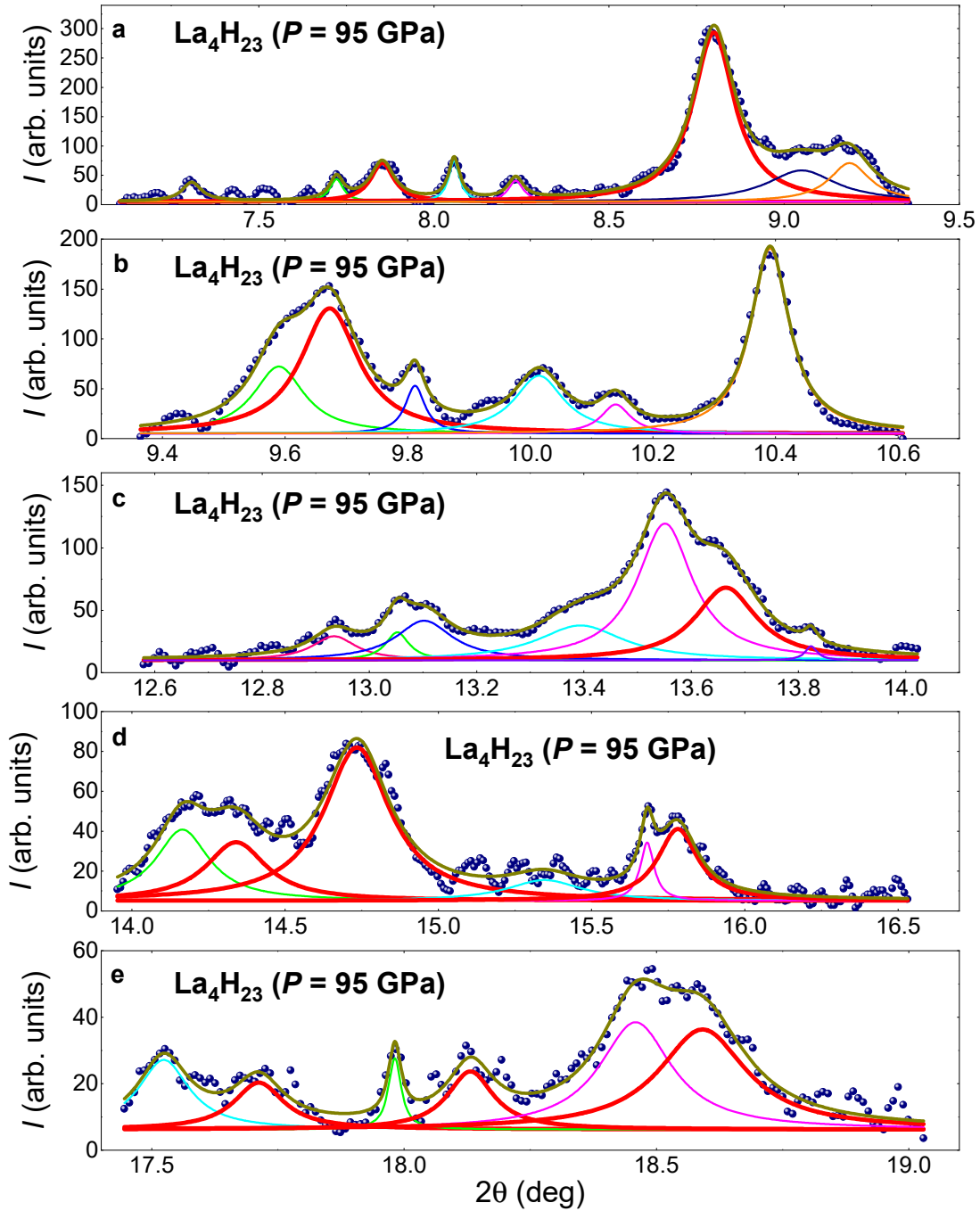

**Figure S1.** XRD scans and data fits to multiple Lorentz peaks function for raw data reported by Cross *et al*[68] for La$_4$H$_{23}$ compressed at $P$ = 95 GPa.

## II. XRD scans and data fit to multiple Lorenz function for data reported by Guo *et al*[67]

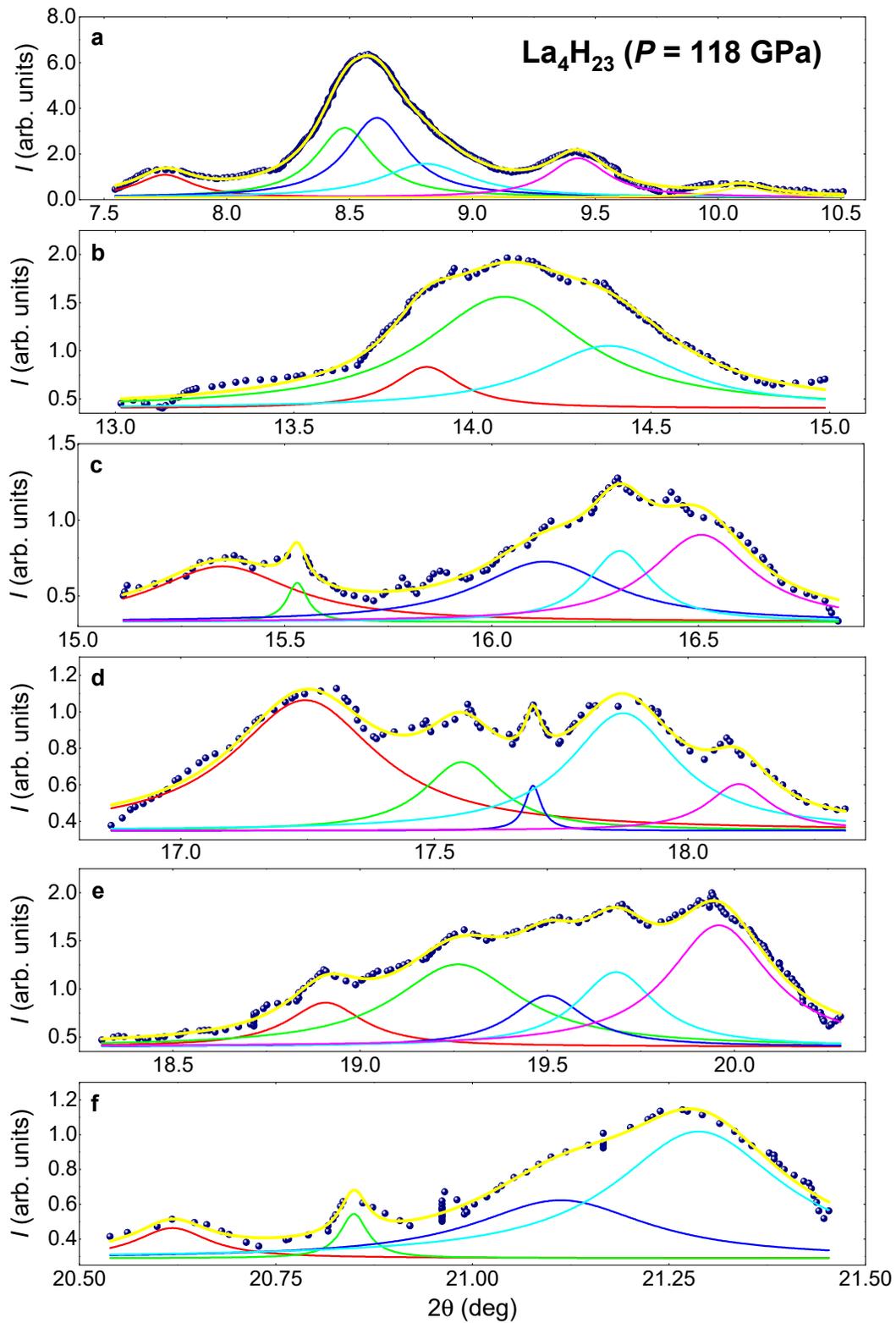

**Figure S2.** XRD scans and data fits to multiple Lorentz peaks function for raw data reported by Guo *et al* [42] for La$_4$H$_{23}$ compressed at $P$ = 118 GPa.

## III. Narrowest peaks in the XRD scan reported by Cross *et al*[68].

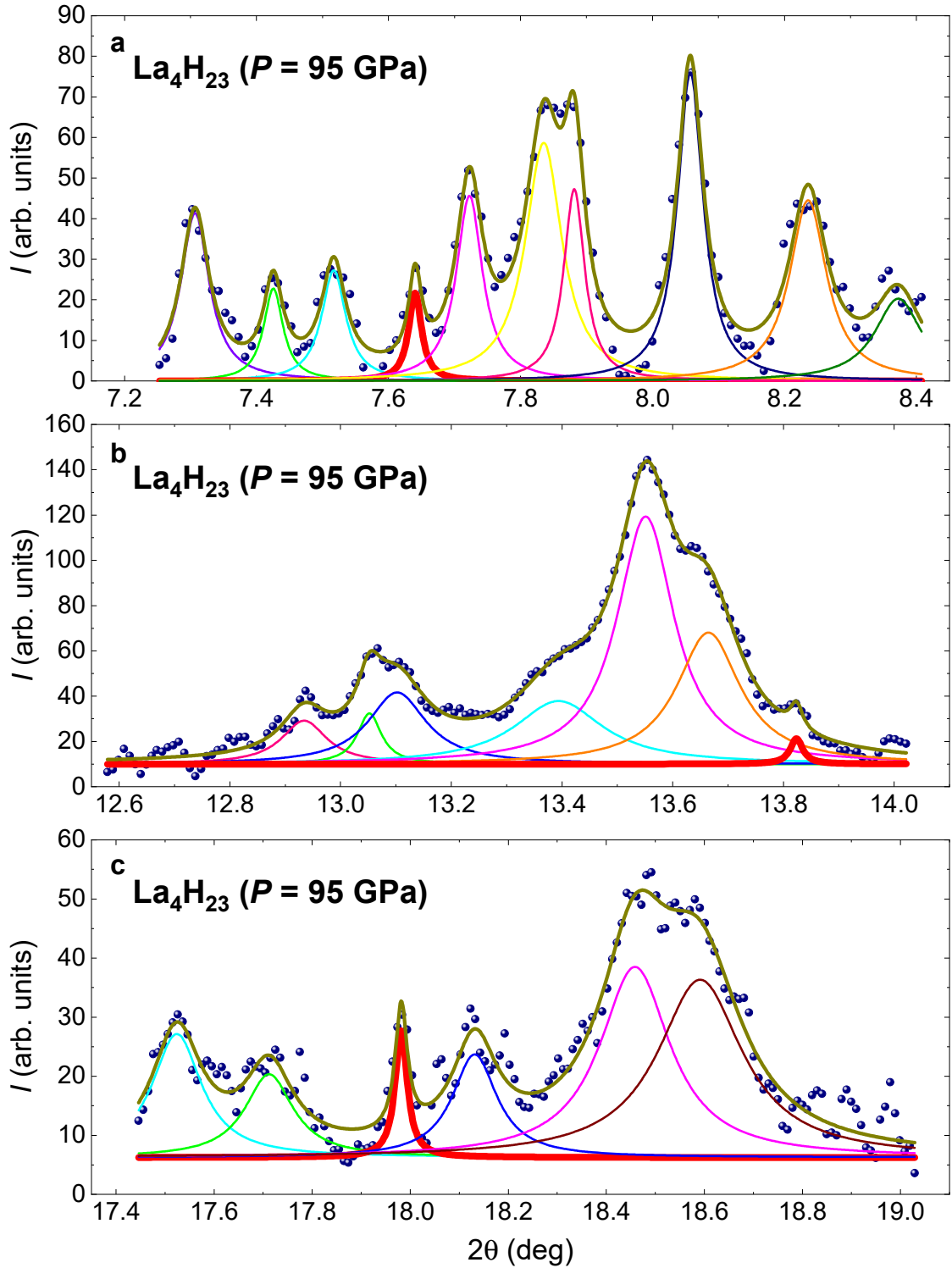

**Figure S3.** Narrowest peaks (thick red lines) in the XRD scan for data reported by Cross *et al*[68] for La$_4$H$_{23}$ compressed at $P$ = 95 GPa.

43